# Comment to "Topological insulators and superconductors" by Xiao-Liang Qi and Shou-Cheng Zhang Reviews of Modern Physics 83, 1057 (2011).

Stanisław Krukowski[1], Paweł Kempisty[1], Paweł Strak[1] and Konrad Sakowski[1,2]

[1]Institute of High Pressure Physics, Polish Academy of Sciences, Sokołowska 29/37, 01-142 Warsaw, Poland

[2]University of Warsaw, Faculty of Mathematics, Informatics and Mechanics, Institute of Applied Mathematics and Mechanics, Banacha 2, 02-097 Warsaw, Poland

In the above review paper (QZ - Qi and Zhang 2011), the authors present derivation of the topological surface state. In Section II they use 2-d model of HgTe half-space $x > 0$, in the x-y plane. The model, referred as BHZ Hamiltonian, published previously by Bernevig et al. (BHZ - Bernevig, Hughes and Zhang 2006) is reduced version of the full eight state HgTe Hamiltonian published by Novik et al. (Novik et al. 2005). The 8 state problem, modified by rejection of the spin-orbit decoupled states, was solved by BHZ as eigenvalue problem for HgTe/CdTe quantum wells (QWs) for several pairs of states separately (BHZ Suppl. online material 2011). Following this example, QZ tackled the half-space (surface) reduced Hamiltonian eigenfunction issue for the envelope function describing the surface state. For small $k$ vector values QZ used the simplified form:

$$\begin{bmatrix} M_k & Ak \\ Ak & -M_k \end{bmatrix} \Psi(x) = E\Psi(x) \qquad (1)$$

where $M_k = M - Bk^2$. Using Peierls substitution ($k = -i\partial_x$) the authors (QZ) solved the eigenvalue problem for zero energy value, i.e. for E = 0. Assuming the envelope function in the form $\Psi(x) = ae^{\lambda x}$ they obtained the space decay exponent as: $\lambda = \mp \left(A \mp \sqrt{A^2 - 4MB}\right)/2B$. The obtained exponent has nonzero real part, for the wavefunction decaying into the interior therefore it was identified as the topological surface state.

The above result arises due to incorrect interpretation of the dispersion relations around Γ point. The solution of the eigenvalue equation (Eq. 1) is:

$$E_{\mp} = \mp\sqrt{(M - Bk^2)^2 + A^2 k^2} \qquad (2)$$

The range of the allowed energy values is: $|E_\mp| \geq \sqrt{\frac{A^2 M}{A^2+2MB}}$, that excludes zero energy value. **Precisely, the use of the zero energy value, outside the allowed energy range, is the source of the error in QZ solution.** The solutions outside the allowed energy range generate spurious surface states.

In fact, HgTe is direct zero bandgap semiconductor as confirmed by high *precision ab initio* calculations, employing hybrid quasi-particle self-consistent Green's function, screened Coulomb interaction (QSGW) scheme (Svane et al. 2011). Svane et al. found that the band structure HgTe is inverted with small $E_g = 0.09 eV$ bandgap. Thus the proper eigenvalue equation is:

$$\begin{bmatrix} M + Bk^2 & Ak \\ Ak & -M - Bk^2 \end{bmatrix} \Psi(x) = E\Psi(x) \tag{3}$$

which gives, for the electron/hole branches, the following dispersion relation:

$$E_\mp = \mp\sqrt{(M + Bk^2)^2 + A^2 k^2} \tag{4}$$

Note that the square root argument is always positive as the sum of the two square numbers. That confirms exclusive existence of bulk modes.

No surface mode exists as follows from the interpretation that can be exemplified by setting the Kane momentum matrix element (Kane 1957) to zero: $P = Ak = 0$, i.e. $A = 0$. For this choice, the solutions (Eq. 4) separate into simple electron and hole dispersion relations:

i/ electrons:

$$E_+ = M + Bk^2 \tag{5a}$$

i/ holes:

$$E_- = -M - Bk^2 \tag{5b}$$

These relations describe bulk solutions having real wavevector k: the electron branch for $E_+ \geq M$ and the hole branch for $E_- \leq -M$. No imaginary wavevector exists, i.e. $\lambda = ik$ is not purely imaginary, excluding solution given by QZ, i.e. the surface state. Note that zero energy is excluded, and accordingly QZ state. The same holds for dispersion relations in Eqs. 2 and 4.

Generally, the electron branch is characterized by positive square k dependence in the surrounding of the energy minimum, i.e.

$$E_e = E_{kin} + V_o = Fk^2 + V_o \tag{6}$$

where $V_o = M$ and $F = B + A^2/2M > 0$ denote the reference potential level and the effective mass coefficients, obtained from Eq. (4). The surface state, described by the real exponent $\lambda = ik = \sqrt{(E_e - V_o)/F}$ may exist for $E_e - V_o = E_{kin} < 0$ only. That is not possible as the kinetic energy is positive definite. Hence the surface state may exist for the potential energy position dependent, i.e. $V_o = V_o(r)$, so that the exponential decay in external region, giving rise to negative contribution to kinetic energy is compensated by the contribution from the low potential energy region where the kinetic energy is positive. Thus existence of the surface state without localization potential is not allowed. Similar conclusions may be reached for the hole branch, given by:

$$E_h = E_{kin} + V_o = -Fk^2 + V_o \qquad (7)$$

where $V_o = -M$ and $F = B + A^2/2M > 0$. Therefore the existence of topological QZ surface state is not compatible with the kinetic energy definition and accordingly it is not possible.

**References:**

Qi X.-L., S.-C. Zhang, 2011, Rev. Mod. Phys. **83**, 1057.

Bernevig B. A., T. L. Hughes, S.-C. Zhang, 2011, Science, **314**, 1757.

Novik E.G, A. Pfeuffer-Jeschke, T. Jungwirth, V. Latoussek, C.R. Becker, G. Landwehr, H. Buchmann, and L.W. Molenkamp, 2005, Phys. Rev. B **72**, 035321.

Svane A, N.E. Christensen, M. Cardona, A.N. Chantis, M. van Schilfgaarde, T. Kotani, 2011, Phys. Rev. B, **84**, 205205.

Kane E.O., J. Phys. Chem. Solids, 1957, **1**, 249.